# Brain Tumor Classification using Vision Transformer with Selective Cross-Attention Mechanism and Feature Calibration

Mohammad Ali Labbaf Khaniki*, Marzieh Mirzaeibonehkhater**, Mohammad Manthouri***, Elham Hasani****

*Faculty of Electrical Engineering, K.N. Toosi University of Technology, Tehran, Iran

** Department of Electrical and Computer Engineering, Indiana University-Purdue University

*** Faculty of Electrical and Electronic Engineering, Shahed university, Tehran, Iran

**** Department of Chemical and Environmental Engineering, University of Cincinnati, Cincinnati, OH 45221, United States

mohamad95labafkh@gmail.com*

marzieh89mirzaei@gmail.com**

mmanthouri@shahed.ac.ir ***

hasaniem@mail.uc.edu****

**Abstract:**

Brain tumor classification is a challenging task in medical image analysis, with significant implications for patient diagnosis and treatment. The objective of this paper is to propose a novel approach to brain tumor classification using a Vision Transformer (ViT) with a novel cross-attention mechanism. Our approach leverages the strengths of transformers in modeling long-range dependencies and multi-scale feature fusion. We introduce two new mechanisms to improve the performance of the cross-attention fusion module: Feature Calibration Mechanism (FCM) and Selective Cross-Attention (SCA). FCM calibrates the features from different branches to make them more compatible, while SCA selectively attends to the most informative features. Our experimental results demonstrate that the proposed approach outperforms other state-of-the-art methods, including Convolutional Neural Networks (CNN), ViT, and Cross-attention ViT, achieving an accuracy of 98.93% and an F1-score of 98.83%. Furthermore, the addition of Stochastic Depth mechanism improves the accuracy to 99.24% and the F1-score to 99.23%. The proposed FCM and SCA mechanisms can be easily integrated into other ViT architectures, making them a promising direction for future research in medical image analysis.

## 1) Introduction

Brain tumors are one of the most common and deadly types of cancer, resulting in a significant number of deaths globally each year. According to the World Health Organization (WHO), brain tumors are responsible for approximately 2% of all cancer-related deaths worldwide [1]. The diagnosis of a brain tumor can have a devastating impact on patients, their families, and the healthcare system as a whole [2]. Brain tumor patients often suffer from severe physical, emotional, and cognitive decline, while their families face emotional strain, financial hardship, and caregiving challenges. Meanwhile, healthcare systems struggle to provide sufficient resources, including specialized care and personnel, to meet the complex needs of these patients [3]. Deep learning, a subset of artificial intelligence, has revolutionized the field of medical image analysis by leveraging large datasets and advanced algorithms to identify patterns and features in medical images, enabling accurate and automated diagnosis [4]. Deep learning offers improved accuracy in brain tumor classification, reducing the risk of misdiagnosis and improving patient outcomes [5]. Additionally, it increases efficiency by significantly reducing the time required for diagnosis, enabling healthcare professionals to focus on more critical tasks [6]. Furthermore, deep learning provides enhanced consistency and enables personalized medicine by identifying specific tumor characteristics, leading to tailored treatment plans and improved patient care.

The Transformer architecture is a type of neural network that has been widely adopted in natural language processing (NLP) tasks. It was introduced in a research paper by Vaswani et al. in 2017 [7]. The Transformer architecture is based on self-attention mechanisms, which allow it to capture dependencies between different parts of the input sequence. Inspired by the success of Transformers in NLP, researchers adapted the architecture for computer vision tasks, giving rise to Vision Transformers (ViTs) [8]. The key innovation of ViTs is to treat fixed-sized image patches as "words" in NLP, allowing the Transformer architecture to be applied to image classification tasks. In ViTs, the input image is divided into fixed-sized patches, which are then fed into the Transformer architecture. The self-attention mechanisms in the Transformer architecture allow it to capture dependencies between different patches in the image, enabling it to learn rich and expressive representations of the image. ViTs offer several advantages over traditional computer vision architectures, including improved performance, flexibility, and interpretability [9]. Additionally, ViTs can be easily adapted to different image classification tasks, making them a versatile and powerful tool for computer vision [10].

The attention mechanism is a powerful technique used in deep learning models, including ViTs, that allows the model to selectively focus on specific regions or features of an image [11], [12]. This allows the model to extract more meaningful representations and improve its performance in various image processing tasks [13]. By prioritizing the most relevant information, attention enables the model to concentrate on the most informative parts of the image, leading to improved accuracy, efficiency, and interpretability. In the context of image classification, attention-based ViTs have demonstrated remarkable effectiveness. By highlighting the most discriminative regions of an image, attention enables ViTs to identify the most relevant features, leading to improved classification accuracy. One of the things that improve the performance of the attention mechanism is the cross-attention mechanism, which enables the integration of information from multiple sources, leading to more comprehensive and accurate representations [14]. One of the benefits of cross-attention is that it facilitates the modeling of complex relationships between different objects or regions in an image, allowing for more sophisticated scene understanding and analysis [15] and [16].

This paper presents a novel and innovative approach to the classification of brain tumors, utilizing advanced techniques and methodologies to improve the accuracy and efficiency of diagnosis. The proposed algorithm differs from existing methods in its use of a novel cross-attention mechanism and three new mechanisms that can be seamlessly integrated into other ViT architectures for brain tumor classification. The dataset used in this research is the Brain MRI Tumor Dataset, which is a comprehensive collection of 3,064 magnetic resonance imaging (MRI) scans of brains, with a focus on the presence or absence of tumors. The dataset is evenly distributed, consisting of 1,532 images labeled as 'Tumor' and 1,532 labeled as 'Non-tumor' [17]. Here are the key advancements of this research:

- **The Feature Calibration Mechanism (FCM):** FCM calibrates the features from the L-branch and S-branch before they are processed by the Cross-Attention module, making them more compatible with each other. The calibration function can be a simple scaling operation, a more complex non-linear transformation, or even a mini neural network. By doing so, FCM offers two key advantages: compatibility, which enables features from different branches to work together seamlessly, particularly when they have different characteristics due to varying scales, modalities, or other factors; and flexibility, which allows the calibration function to be tailored to specific data and task requirements, making it adaptable to diverse scenarios.
- **The Selective Cross-Attention (SCA) Mechanism:** SCA selectively attends to the most informative patch tokens from the S-branch based on their relevance to the CLS token from the L-branch, calculating a relevance score for each patch token and performing the Cross-Attention operation only on a subset of the most relevant tokens. This approach offers two key advantages: efficiency, as focusing on the most informative features can lead to faster training and inference times; and effectiveness, as SCA enables the model to ignore less relevant features, potentially reducing noise and improving the overall performance of the Cross-Attention Fusion module.

- **The Stochastic Depth Mechanism:** This mechanism randomly drops out layers during training, which helps to improve the robustness and generalizability of the model. By randomly dropping out layers, the model is forced to learn multiple representations of the input data, making it more robust to variations in the input data. This approach also helps to prevent overfitting, as the model is not able to rely on any single layer or representation.

The general results of this research are highly promising, with the proposed approach outperforming existing methods by achieving an accuracy of 98.93% and an F1-score of 98.83%, thereby surpassing the performance of the Convolutional Neural Networks (CNN) [4], RanMerFormer [18], vViT [19], ViT, and Cross ViT. Furthermore, the addition of the Stochastic Depth mechanism takes the performance to new heights, with an accuracy of 99.24% and an F1-score of 99.23%. Overall, the proposed approach demonstrates its effectiveness in improving the accuracy and efficiency of brain tumor classification, outperforming state-of-the-art results in this field.

The manuscript is organized as follows. The second section provides an overview of the brain MRI tumor dataset used in the study. The third section describes the proposed methodology, which FCM and SCA mechanisms within the ViT framework. IV presents the outcome of the simulations, providing a thorough account of the training and validation processes used to assess the proposed approach. The fifth section wraps up the study by summarizing the key results and emphasizing the significant contributions of this research.

## 2) Background and Related Work

This section provides a comprehensive literature review of recent advancements in machine learning and deep learning, with a focus on their applications in medical imaging and brain tumor detection. Traditional CNN-based approaches have limitations, but novel methods like ViT-based models have shown promise in tumor detection by leveraging self-attention mechanisms. This section also provides an overview of the Brain MRI Tumor Dataset, a comprehensive collection of 3,064 MRI scans, which is publicly available for research purposes and serves as a valuable resource for developing and testing algorithms for brain tumor detection and classification.

## 2-1) Brain MRI Tumor Dataset Overview

The Brain MRI Tumor Dataset is a comprehensive collection of 3,064 MRI scans, specifically of brains, with a focus on the presence or absence of tumors. The dataset is evenly distributed with 1,532 images labeled as 'Tumor' and 1,532 labeled as 'Non-tumor'. Each image is of the resolution 256x256 pixels and is in the DICOM (Digital Imaging and Communications in Medicine) format. This dataset has been meticulously curated from various hospitals and medical research institutions, ensuring a diverse range of image characteristics and tumor types [17]. The MRI scans were acquired using different scanners and protocols, adding to the diversity of the dataset. Each image in the dataset is annotated with a binary label indicating the presence (1) or absence (0) of a tumor. These annotations were performed by experienced radiologists and medical experts, ensuring the reliability of the labels. The dataset has undergone rigorous quality control. All images were reviewed for artifacts, noise, and other quality issues. Any images that did not meet the quality standards were removed from the dataset.

The Brain MRI Tumor Dataset is publicly available for research purposes. It is designed to facilitate the development and evaluation of machine learning and deep learning models for brain tumor detection, segmentation, and classification. The dataset can be downloaded from Kaggle. This dataset serves as a valuable resource for researchers and practitioners in the field of medical imaging, machine learning, and healthcare. It provides a robust platform for developing and testing algorithms for brain tumor detection and classification. The diverse range of image characteristics and tumor types in the dataset ensures that models trained on this dataset can generalize well to unseen data. The meticulous curation and annotation of the dataset ensure its reliability and usefulness in research and application. To provide a visual understanding of the classification task, sample images from both classes are presented below.

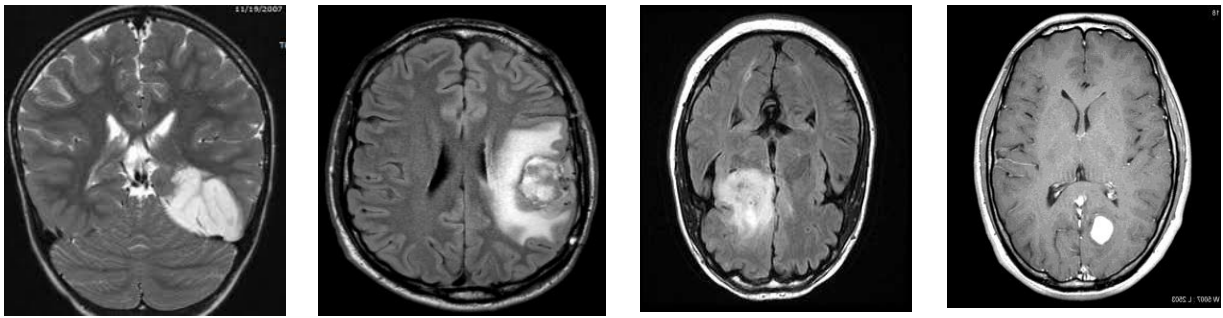

(a)

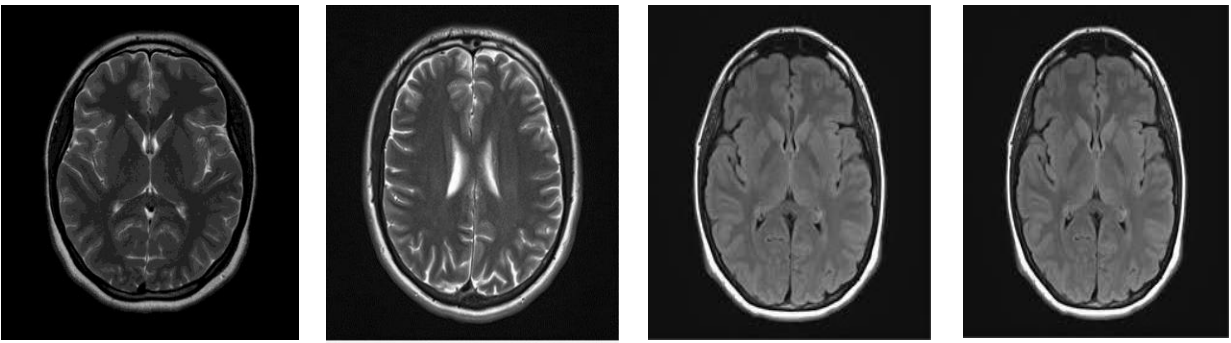

(b)

Figure 1: Representative examples of (a) Brain Tumor MRI and (b) Healthy Brain MRI Scans.

## 2-2) Literature review

Machine Learning, a branch of Artificial Intelligence, enables computers to autonomously learn from data, adapt, and refine their performance over time with minimal human intervention. This technology can be applied to a wide range of tasks, such as developing including multi-objective optimization [20], conversational agents [21], and photovoltaic prediction performance [22], and virtual reality [23] and [24]. Deep Learning (DL), a sophisticated branch of ML, replicates the human brain's learning mechanism through artificial neural networks. Its applications have sparked widespread interest across diverse fields, including the detection of straggler tasks [25], real-time motion pattern analysis [26], and road monitoring [27]. [28] utilizes deep learning techniques to predict financial market sequences, aiming to provide valuable insights for policymakers to enhance economic policies and make informed decisions. [29] and [30] present the Grace Platform, an innovative educational tool that leverages gamified augmented reality and virtual reality to enhance pedagogy and improve learning outcomes in agriculture education, exploring the potential of immersive technologies to gamify learning and enhance agricultural education through interactive and engaging experiences. [31] introduces a novel approach to pre-training LiDAR perception models using masked autoencoders, enabling ultra-efficient 3D sensing by generating high-quality point clouds from sparse inputs. [32] proposes a blockchain-based scheme for secure and decentralized medical data sharing with searchable encryption, empowering healthcare with confidentiality and efficiency.

Traditional deep learning approaches in medical imaging rely on CNNs that are limited by their inability to model long-range dependencies and capture complex contextual relationships in images, leading to suboptimal performance in certain tasks. [33] presents a brain tumor classification approach using deep CNN features via transfer learning, achieving high accuracy and robustness in identifying tumor types from medical images. [34] evaluates the performance of Foundational Medical's Segment Anything (Med-SAM1, Med-SAM2) deep learning models in segmenting the left atrium in 3D MRI. [35] uses gray level enhancement and deep learning to accurately classify white matter lesions in multiple sclerosis patients' MRI images. [36] proposes a deep learning-based approach to predict 3D human motion trajectories in construction settings, enabling proactive safety measures and improved worker productivity through activity-aware motion forecasting. [37] proposes the use of various artificial intelligence methods to detect COVID-19 enabling accurate and rapid diagnosis of the disease. [38] proposes an automatic brain tumor detection system using a CNN transfer learning approach, enabling accurate and efficient identification of tumors from medical images. [39] presents a novel approach to robot-assisted physiotherapy, using dynamic time warping and recurrent neural networks to enable multi-joint adaptive motion imitation. [40] presents a CNN-based approach for human brain tumor classification and segmentation, achieving accurate detection and delineation of tumors from medical images.

ViT-based models, a novel deep learning approach in medical imaging, have shown remarkable potential in tumor detection by harnessing the power of self-attention mechanisms to process images as sequences of patches, enabling more accurate and robust detection of tumors.[41] proposes a hybrid approach combining transformer and CNN for effective brain tumor classification using MRI images, achieving improved accuracy and robustness in tumor diagnosis. [42] introduces LCDEiT, a linear complexity data-efficient image transformer, which achieves state-of-the-art performance in MRI brain tumor classification. [43] proposes a transformer-based multi-style network for human pose prediction in industrial human-robot collaboration, leveraging a custom data pipeline to improve accuracy. [44] presents BTSwin-Unet, a 3D U-shaped symmetrical Swin transformer-based network for brain tumor segmentation, which leverages self-supervised pre-training to achieve accurate and robust tumor segmentation from MRI images. [45] proposes CKD-TransBTS, a clinical knowledge-driven hybrid transformer that incorporates modality-correlated cross-attention, achieving state-of-the-art performance in brain tumor segmentation by effectively integrating multi-modal MRI data and clinical knowledge. [46] introduces the Focal Cross Transformer, a multi-view brain tumor segmentation model that leverages cross-window and focal self-attention mechanisms to capture complex tumor structures and relationships, achieving improved segmentation accuracy and robustness.

Traditional DL approaches in medical imaging, including CNN or even ViT, have several limitations. This paper proposes a novel method that addresses these shortcomings:

- **Inefficient Cross-Attention**: Traditional ViT models perform Cross-Attention on all patch tokens, leading to computational inefficiency and reduced effectiveness. Our proposed method introduces the Selective Cross-Attention (SCA) Mechanism, which selectively attends to the most informative patch tokens, improving efficiency and effectiveness.

- **Limited Robustness and Generalizability**: Traditional ViT models may not be robust to input variations and may overfit to training data. Our proposed method incorporates the Stochastic Depth Mechanism, which randomly drops out layers during training, enhancing robustness and generalizability.

- **Insufficient Handling of Multi-Scale Features**: Traditional ViT models may not effectively handle multi-scale features, common in medical imaging data. Our proposed method includes the L-branch and S-branch, which can handle features at different scales and modalities.

- **Lack of Flexibility:** Traditional ViT models may not be adaptable to different data and task requirements. Our proposed method includes the calibration function in the Feature Calibration Mechanism (FCM), which can be tailored to specific data and task requirements, making it adaptable to diverse scenarios.

By addressing these limitations, our proposed method aims to improve the performance and robustness of DL approaches in medical imaging.

## 3) Methodology

In this section, we introduce the Cross-Attention Mechanism specifically designed for ViT. We then delve into the details of the Feature Calibration Mechanism, Selective Cross-Attention, and Stochastic Depth Mechanism, providing an in-depth explanation of the underlying intuition behind each formula.

### 3-1) Cross-Attention Mechanism

The Cross-Attention Fusion module is designed to amalgamate features from the large branch (L-branch) and small branch (S-branch). The module comprises the following components:

**Token Concatenation**: The Class (CLS) token from the L-branch and the patch tokens from the S-branch are amalgamated to form a novel token sequence, denoted as $\mathbf{x}'^l$. This can be mathematically represented as:

$$\mathbf{x}'^l = \left[ f^l(\mathbf{x}_{cls}^l) \mid \mathbf{x}_{patch}^s \right] \tag{1}$$

Here, $f^l(\cdot)$ is the projection function employed to align the dimensions. Equ. (1) project the CLS token from the L-branch to match the dimensionality of the patch tokens from the S-branch, and then merge them into a single token sequence that integrates both global and local information, enabling the model to capture long-range and short-range dependencies in the input data.

**Cross-Attention (CA)**: The CA module accepts the concatenated token sequence $\mathbf{x}'^l$ as input and executes attention between the CLS token and the patch tokens. The CLS token functions as the query, while the patch tokens act as the key and value. This can be expressed as:

$$\mathbf{q} = \mathbf{x}'^l_{cls}\mathbf{W}_q, \mathbf{k} = \mathbf{x}'^l\mathbf{W}_k, \mathbf{v} = \mathbf{x}'^l\mathbf{W}_v \tag{2}$$

$$\mathbf{A} = softmax\left(\frac{\mathbf{q}\mathbf{k}^T}{\sqrt{\frac{C}{h}}}\right) \tag{3}$$

$$CA(\mathbf{x}'^l) = \mathbf{A}\mathbf{v} \tag{4}$$

In these equations, $\mathbf{W}_q$, $\mathbf{W}_k$, and $\mathbf{W}_v$ are learnable parameters, $C$ represents the embedding dimension, and $h$ is the number of heads. In (3), The attention weights represent the importance of each patch token in relation to the CLS token. Equ. (4) represents the features of the patch tokens that are most relevant to the CLS token.

**Multi-Head Cross-Attention (MCA)**: The CA module is extended to multiple heads, analogous to self-attention. Given the query matrix $Q$, key matrix $K$, and value matrix $V$, each of which is obtained by multiplying the input matrix $X$ with the corresponding weight matrix (i.e. $Q = XW^q$, $K = XW^k$, $V = XW^v$ ), the output of a single attention head can be calculated as:

$$\text{head}_i = \text{softmax}\left(\frac{Q_i K_i^T}{\sqrt{d_k}}\right) V_i \tag{5}$$

where $d_k$ is the dimensionality of the keys, which is typically set to $d_{\text{model}}/h$, and $h$ is the number of heads. Equ. (5) computes attention between query and key matrices, outputting a weighted sum of the value matrix. The outputs of all the attention heads are then concatenated and linearly transformed to produce the final output:

$$\text{MCA}(X) = \text{Concat}(\text{head}_1, \text{head}_2, \ldots, \text{head}_h) W^o \tag{6}$$

Where $W^o$ is a learnable weight matrix. In (6), Concatenates outputs of all attention heads and applies a linear transformation.

**Layer Normalization and Residual Shortcut**: The output of the MCA module is added to the original CLS token, followed by layer normalization and a residual shortcut. This can be represented as:

$$\mathbf{y}_{cls}^l = f^l(\mathbf{x}_{cls}^l) + MCA(LN([f^l(\mathbf{x}_{cls}^l) \| \mathbf{x}_{patch}^S])) \tag{7}$$

$$\mathbf{z}^l = [g^l(\mathbf{y}_{cls}^l) \| \mathbf{x}_{patch}^l] \tag{8}$$

Here, $f^l(\cdot)$ and $g^l(\cdot)$ are the projection and back-projection functions for dimension alignment, respectively. Equ. (7) Refine the original CLS token by incorporating information from the patch tokens through the MCA mechanism, and then combine the results to get an enhanced representation. Equ. (8) Transform the refined CLS token to match the dimensionality of the patch tokens, and then merge them together to produce the final output.

The Cross-Attention Fusion module is more efficient than the All-Attention Fusion module as it solely utilizes the CLS token as the query, thereby reducing the computational and memory complexity of generating the attention map. Empirical evidence demonstrates that Cross-Attention outperforms other simple heuristic approaches in terms of accuracy, while also being efficient for multi-scale feature fusion. Cross-Attention Fusion in ViT diagram has been shown in Fig. 2.

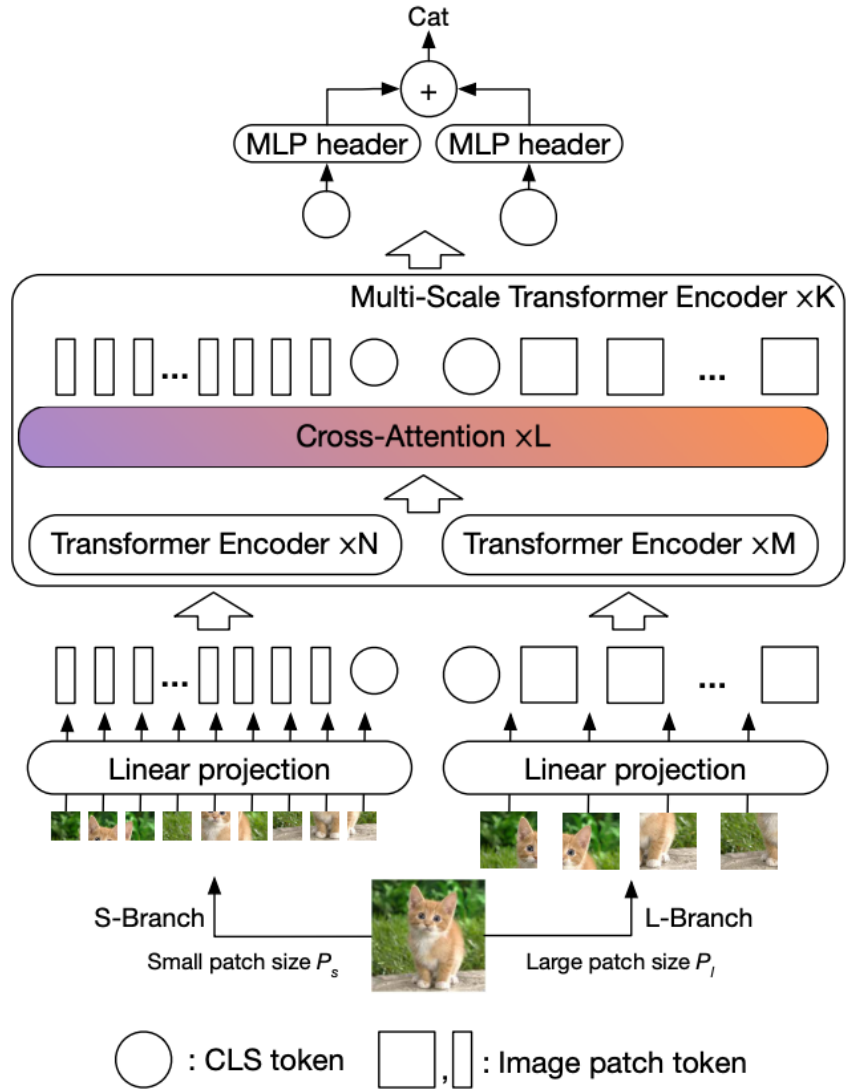

Figure 2: Cross-Attention Fusion in (ViT) Architecture [47].

## 3-2) Feature Calibration Mechanism and Selective Cross-Attention

### A. Feature Calibration Mechanism (FCM)

Before the token concatenation step, the features from the L-branch and S-branch are calibrated using a calibration function $c(\cdot)$. This function could be a simple scaling operation, a more complex non-linear transformation, or even a mini neural network. The calibrated features are denoted as $x_{cal}^l$ and $x_{cal}^s$. Formally, we have:

$$x_{cal}^l = c(x^l) \qquad (9)$$

$$x_{cal}^s = c(x^s) \qquad (10)$$

Equ. (9-10) calibrates features from L-branch and S-branch using a calibration function, making them compatible. The calibrated CLS token from the L-branch and the calibrated patch tokens from the S-branch are concatenated to form a new token sequence $x_{cal'}^l$. Formally, we have:

$$x_{cal'}^l = [f_l(x_{cal}, cls_l) \parallel x_{cal,patch}^s] \qquad (11)$$

Equ. (11) is taking the calibrated features from both branches and "stitching" them together to create a new, combined feature representation. This combined representation can then be used as input to the Cross-Attention mechanism, allowing the model to attend to the most relevant features from both branches.

## B. Selective Cross-Attention (SCA) Mechanism

Before the Cross-Attention operation, a relevance score is calculated for each patch token from the S-branch. This score measures the relevance of the patch token to the CLS token, and can be calculated using a small neural network or a simple dot product operation. Formally, we have:

$$r = RelevanceScoreCalculator(x_{cls}^l, x_{patch}^s) \qquad (12)$$

The Cross-Attention operation calculates a relevance score for each patch token from the S-branch, measuring its importance to the CLS token, and then selectively apply Cross-Attention to the top-scoring tokens, allowing the model to prioritize the most informative features and filter out less relevant information. Formally, we have:

$$x_{sel'}^l = SelectTopK(x_{cal'}^l, r, K) \qquad (13)$$

$$q = x_{sel,cls'}^l W^q, k = x_{sel'}^l W^k, v = x_{sel'}^l W^v \qquad (14)$$

$$A = softmax\left(\frac{qk^T}{\sqrt{\frac{C}{h}}}\right) \tag{15}$$

$$SCA(x_{sel'}^l) = Av \tag{16}$$

The rest of the steps (Multi-Head Cross-Attention, Layer Normalization and Residual Shortcut) remain the same as in the original Cross-Attention Fusion module, but they operate on the selectively attended and calibrated features $x_{sel'}^l$ and $x_{cal'}^l$. The SCA mechanism is designed to selectively focus on the most informative features by computing a relevance score for each patch token and selecting the top-scoring tokens. The mechanism then transforms the selected tokens into query, key, and value vectors, computes attention weights, and takes a weighted sum of the value vectors to produce the output. This allows the model to prioritize the most relevant features and ignore less relevant information, enabling more efficient and effective processing of the input data. The rest of the steps (Multi-Head Attention and Layer Normalization and Residual Shortcut) remain the same as in the original Cross-Attention Fusion module, but they operate on the calibrated features

### C. Stochastic Depth Mechanism

The Stochastic Depth Mechanism is another key advancement introduced in this research. This mechanism randomly drops out layers during the training process, which serves two main purposes:

- Improving Robustness: By randomly dropping out layers, the model is forced to learn multiple representations of the input data. This makes the model more robust to variations in the input data, enhancing its ability to generalize from the training data to unseen data.

- Preventing Overfitting: The stochastic depth mechanism also helps to prevent overfitting. Overfitting occurs when a model learns the training data too well, to the point where it performs poorly on unseen data. By not allowing the model to rely on any single layer or representation, the stochastic depth mechanism ensures that the model generalizes well.
- Enhance feature fusion: By randomly dropping out layers, the model is encouraged to fuse features from different branches in a more adaptive and flexible manner.

The stochastic depth mechanism can be mathematically represented as follows:

Let $L$ be the total number of layers in the model. For each training step, a layer $l$ is randomly selected with probability $p_l$, where $p_l = 1 - \frac{l}{L}$, and dropped out of the model. The output of the model is then computed based on the remaining layers.

### 4) Simulations

This section presents a comprehensive evaluation of the performance of several deep learning models, including ViT, Cross ViT, Proposed Cross ViT, and Proposed Cross ViT with Stochastic Depth.

The SHAP (SHapley Additive exPlanations) method is a model-agnostic technique that assigns a value to each feature for a specific prediction, providing a quantitative measure of the contribution of each feature to the model's output. Insights into the relationships between hyperparameters and model performance can be gained by leveraging SHAP, facilitating a deeper understanding of the ViT model's behavior. Figure 3 shows the best hyperparameters for the ViT model, achieving the highest validation accuracy.

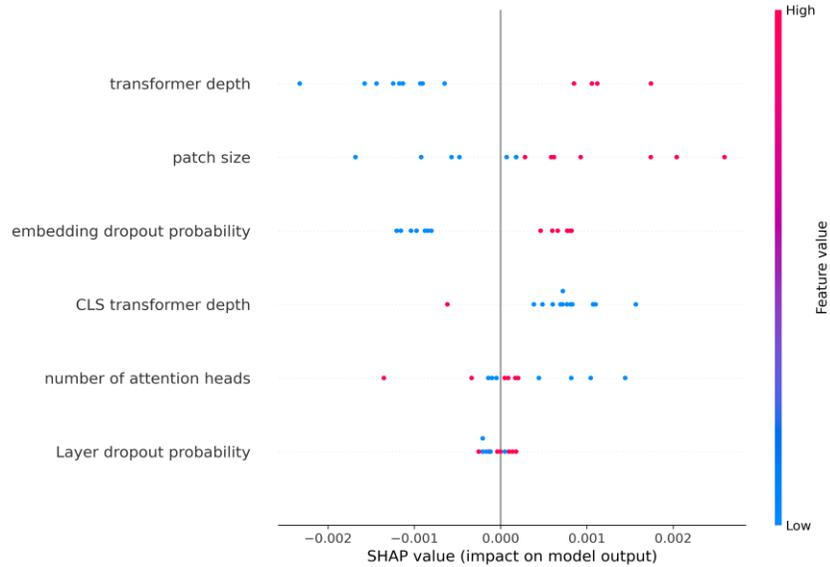

Figure 3: SHAP summary plot of hyperparameter importance for ViT model performance.

According to Fig.3, the main hyperparameters for the Proposed Cross ViT with Stochastic Depth are transformer depth, patch size, embedding dropout probability, CLS transformer depth, number of attention heads, and layer dropout probability, which play a crucial role in determining the model's performance and behavior. Notably, transformer depth and patch size are the most significant factors, having the greatest impact on the model's performance.

The evaluation is based on traditional metrics, and the models are simulated using the following hyperparameters:

- Image size: $256 \times 256$ pixels
- Patch size: $16 \times 16$ pixels
- Number of classes: 2 (binary classification)
- Dimensionality: 256 (dimensionality of the patch embeddings)
- Transformer depth: 6 (depth of the transformer for patch-to-patch attention only)

- CLS transformer depth: 2 (depth of the cross-attention of CLS tokens to patch)
- Number of attention heads: 12
- MLP dimension: 512 (dimensionality of the multilayer perceptron)
- Dropout probability: 0.15 (dropout probability for the transformer layers)
- Embedding dropout probability: 0.15 (dropout probability for the patch embeddings)
- Layer dropout probability: 0.05 (probability of randomly dropping out 5% of the layers during training)

The learning rate and loss function used in the simulations are 0.001 and cross-entropy, respectively.

The simulation results demonstrate the superiority of the proposed method, Proposed Cross ViT with Stochastic Depth, over the other models. The proposed method's performance improvement can be attributed to the incorporation of stochastic depth, which enables the model to adapt to different input patterns and improve its robustness.

### 4.1) Accuracy

This subsection presents the accuracy as an evaluation metric to assess the performance of the models. The accuracy formula is given by following formula.

$$Accuracy = \frac{TP + TN}{TP + TN + FP + FN} \tag{17}$$

In the context of binary classification, TP (True Positives) refers to the number of actual positive instances correctly predicted as positive, TN (True Negatives) refers to the number of actual negative instances correctly predicted as negative, FP (False Positives) refers to the number of actual negative instances incorrectly predicted as positive, and FN (False Negatives) refers to the number of actual positive instances incorrectly predicted as negative. Figure 4 illustrates the accuracy performance over 50 epochs of four different models during training process on test dataset: ViT, Cross ViT, Proposed Cross ViT, and Proposed Cross ViT with Stochastic Depth.

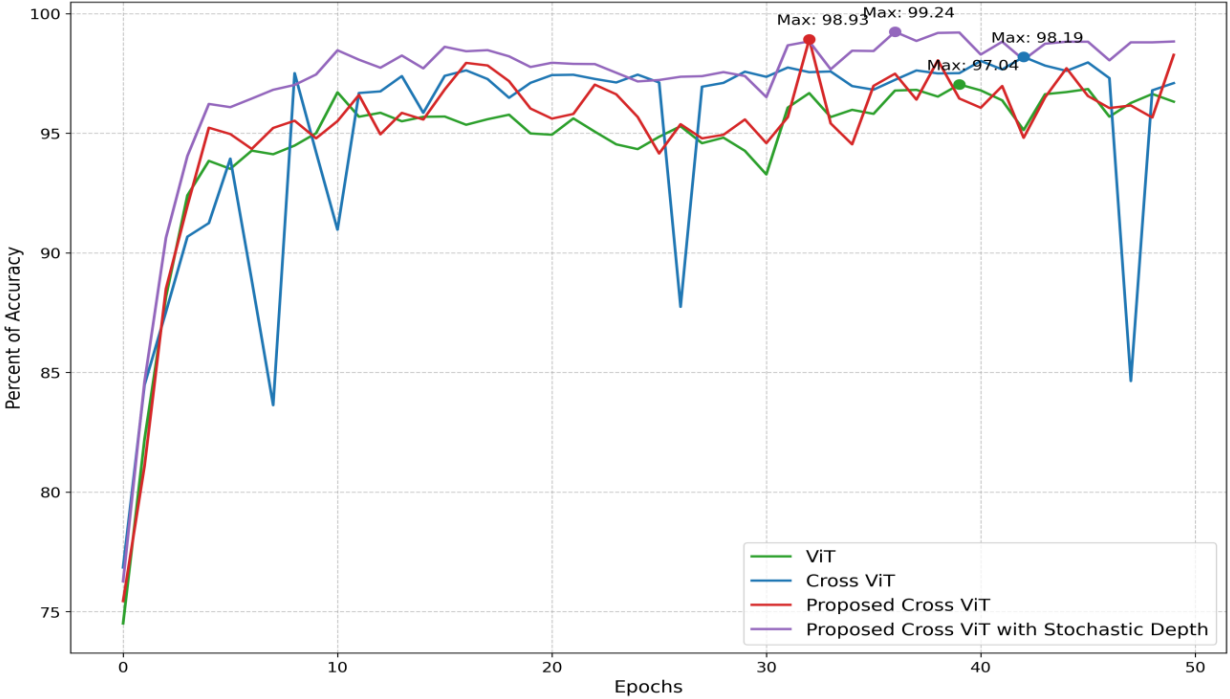

Figure 4: Accuracy curves showing the percentage of accuracy achieved by the ViT, Cross ViT, Proposed Cross ViT, and Proposed Cross ViT with Stochastic Depth models on the test dataset during the training process.

Figure 4 reveals the following insights into the performance of the four models:

- The model exhibits an initial increase in accuracy, followed by fluctuations, ultimately reaching a maximum accuracy of 97.04%.
- Cross ViT: This model displays higher initial variability, but eventually stabilizes, achieving a peak accuracy of 98.19%. However, it experiences significant accuracy drops at certain points, indicating instability.

- Proposed Cross ViT: This model demonstrates a smoother increase in accuracy compared to ViT, reaching a maximum of 98.93%. Its performance is more stable, with fewer fluctuations than Cross ViT. The integration of the Feature Calibration Mechanism (FCM) and Selective Cross-Attention (SCA) Mechanism contributes to its superior performance and stability. FCM enhances feature compatibility and flexibility, while SCA improves efficiency and effectiveness by focusing on the most informative features
- Proposed Cross ViT with Stochastic Depth: This model achieves the highest maximum accuracy of 99.24%, outperforming the other models. It maintains high accuracy with minimal fluctuations throughout the epochs, indicating the effectiveness of stochastic depth in improving model stability and performance.

Overall, the Proposed Cross ViT with Stochastic Depth consistently outperforms the other models in terms of maximum accuracy and stability, suggesting that the addition of stochastic depth is beneficial in enhancing the performance of Cross ViT.

### 4.2) Receiver Operating Characteristic

The Receiver Operating Characteristic (ROC) curve is a graphical representation of the performance of a binary classification model, plotting the True Positive Rate (TPR), also known as Sensitivity or Recall, against the False Positive Rate (FPR), also known as 1-Specificity or Fall-out, at various threshold settings. The ROC curve is a two-dimensional representation of the trade-off between the model's ability to correctly classify TP and its tendency to misclassify FP. The area under the ROC curve (AUC-ROC) is a metric that quantifies the model's performance, with values ranging from 0 (random guessing) to 1 (perfect classification). Figure 5 illustrates the AUC-ROC for test dataset.

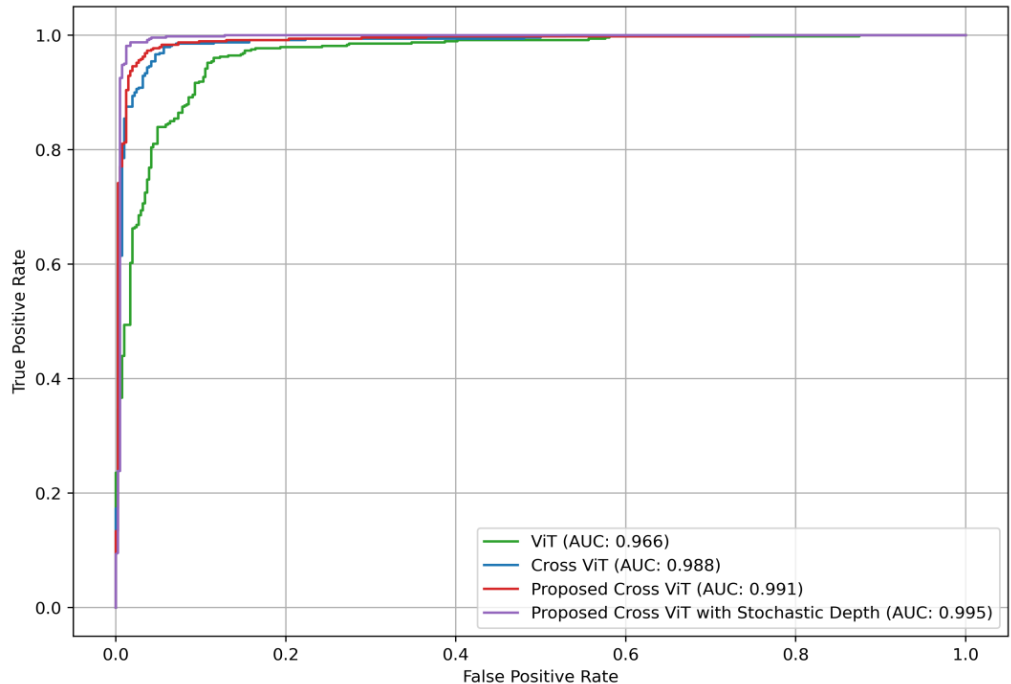

Figure 5: ROC curves showing the AUC for the ViT, Cross ViT, Proposed Cross ViT, and Proposed Cross ViT with Stochastic Depth models on the test dataset.

Figure 5 presents a comparison of the AUC values for the four models, providing insights into their performance in distinguishing between classes.

- ViT: The baseline ViT model performs relatively poorly, with an AUC of 0.966. This suggests that the model may struggle to accurately classify instances, particularly in cases where the classes are closely related.

- Cross ViT: The Cross ViT model performs well, with an AUC of 0.988. This indicates that the model is able to effectively distinguish between classes, with a high degree of accuracy.

- Proposed Cross ViT: The Proposed Cross ViT model shows improved performance, with an AUC of 0.991. This suggests that the integration of (FCM) and (SCA) Mechanism has a positive impact on the model's ability to classify instances accurately.

- Proposed Cross ViT with Stochastic Depth: The Proposed Cross ViT with Stochastic Depth model achieves the highest AUC of 0.995, indicating superior performance in classification tasks among the four models. The addition of stochastic depth appears to further enhance the model's ability to accurately distinguish between classes.

In general, a higher AUC indicates better performance in distinguishing between classes. The Proposed Cross ViT with Stochastic Depth model stands out as the best performer, followed closely by the Proposed Cross ViT.

### 4.3) Quantitative Evaluation of the Classification Methods

This section presents the quantitative evaluation of the classification methods, where the performance of four models is assessed using three key metrics: Recall, Precision, and F1-score. These metrics are calculated using formulas (18), (19), and (20), respectively, and the results are presented in Table 1, providing a detailed understanding of the models' efficacy.

In order to evaluate the effectiveness of our proposed model, we compare its performance with other state-of-the-art deep learning architectures in the field of brain tumor classification, including RanMerFormer [18] and variable Vision Transformer (vViT) [19]. RanMerFormer is a novel deep learning architecture that combines the strengths of ViTs and token merging techniques to achieve state-of-the-art performance in brain tumor classification [18]. By introducing randomness in the token merging process, RanMerFormer improves the model's robustness and ability to generalize to new, unseen data. [19] proposes a novel approach to grading diffuse glioma based on the 2021 WHO classification using a variable Vision Transformer (vViT) architecture, which leverages self-attention mechanisms to effectively capture complex spatial relationships in MRI images. The vViT model demonstrates promising results in accurately grading diffuse glioma, showcasing the potential of deep learning-based approaches in improving diagnostic accuracy and clinical decision-making in neuro-oncology.

$$Recall = \frac{TP}{TP + FN} \tag{18}$$

$$Precision = \frac{TP}{TP + FP}. \tag{19}$$

$$F1-score = 2\frac{Precision \times Recall}{Precision + Recall}. \tag{20}$$

Table 1. Comparison of classification performance metrics (accuracy, recall, precision, and F1-score) for different models, including CNN [4], RanMerFormer [18], vViT [19], ViT, Cross ViT, Proposed Cross ViT, and Proposed Cross ViT with Stochastic Depth.

| Method | Accuracy (%) | Recall (%) | Precision (%) | F1-score (%) |
|---|---|---|---|---|
| CNN [4] | 95.32 | 95.28 | 95.36 | 95.32 |
| RanMerFormer [18] | 99.15 | 99.26 | 99.05 | 99.15 |
| vViT [19] | 99.06 | 98.98 | 99.10 | 99.04 |
| Vit | 97.04 | 97.17 | 96.95 | 97.06 |
| Cross Vit | 98.19 | 98.17 | 98.16 | 98.16 |
| Proposed Cross ViT | 98.93 | 98.95 | 98.72 | 98.83 |
| Proposed Cross ViT with Stochastic Depth | 99.24 | 99.27 | 99.20 | 99.23 |

Table 1 presents a comprehensive comparison of the performance of the six models, including the Proposed Cross ViT with Stochastic Depth, Proposed Cross ViT, Cross Vit, Vit, CNN [4], RanMerFormer [18], and vViT [19]. The results demonstrate that all four models perform well, with high accuracy, recall, precision, and F1-score values.

- The Proposed Cross ViT with Stochastic Depth model outperforms the others, achieving the highest recall (99.27%), precision (99.20%), and F1-score (99.23%).

- The Proposed Cross ViT model is a close second, with a recall of 98.95%, precision of 98.72%, and F1-score of 98.83%.

- The Cross Vit model performs well, with a recall of 98.17%, precision of 98.16%, and F1-score of 98.16%.

- The Vit model has a lower performance, with a recall of 97.17%, precision of 96.95%, and F1-score of 97.06%.

- The CNN [4] model has the lowest performance, with a recall of 95.28%, precision of 95.36%, and F1-score of 95.32%.

- The RanMerFormer [18] model achieves a recall of 99.26%, precision of 99.05%, and F1-score of 99.15%, while the vViT [19] model achieves a recall of 98.98%, precision of 99.10%, and F1-score of 99.04%. Although these models demonstrate high performance, they are outperformed by the Proposed Cross ViT with Stochastic Depth model, which achieves a recall of 99.27%, precision of 99.20%, and F1-score of 99.23%.

The results suggest that the Proposed Cross ViT with Stochastic Depth model is the most effective at classifying the data accurately. The addition of stochastic depth to the Proposed Cross ViT model improves its performance, making it a more robust and accurate model. The Proposed Cross ViT model is a close second, indicating that the integration of the FCM and SCA Mechanism is effective in improving the model's performance.

The Proposed Cross ViT with Stochastic Depth model achieved superior performance, but the study has limitations, including limited dataset evaluation, and lack of hyperparameter investigation which should be addressed in future research.

## 5) Conclusion

In conclusion, this paper presents a novel approach to brain tumor classification using a ViT with a cross-attention mechanism. The proposed model, Cross ViT, incorporates three key advancements: the FCM, SCA, and the Stochastic Depth Mechanism. These mechanisms collectively contribute to the model's superior performance and stability. The FCM and SCA Mechanisms play a crucial role in enhancing the compatibility and flexibility of features from different branches, as well as improving efficiency and effectiveness by focusing on the most informative features. The Stochastic Depth Mechanism further improves the model's robustness and generalizability by randomly dropping out layers during training. The simulation results demonstrate the effectiveness of the proposed model, with the Proposed Cross ViT with Stochastic Depth achieving the highest accuracy of 99.24%, outperforming the other models. Additionally, the model also exhibits superior performance in terms of AUC, with a value of 0.995, indicating its ability to accurately distinguish between classes. Moreover, the results of the other metrics, including Recall, Precision, and F1-score, also demonstrate the proposed model's superior performance. The Proposed Cross ViT with Stochastic Depth achieves the highest values for these metrics, indicating its ability to accurately classify brain tumors. Overall, this paper presents a significant contribution to the field of brain tumor classification, offering a novel and effective approach that can be used in clinical practice. The proposed model's superior performance and stability make it a promising tool for improving the accuracy and efficiency of brain tumor diagnosis.

Future research directions include integrating multiple imaging modalities (MRI, CT, PET) to enhance performance and developing methods to estimate prediction uncertainty, enabling clinicians to make informed decisions and improving diagnostic accuracy and confidence.

## 6) Reference

[1]     J. R. McFaline-Figueroa and E. Q. Lee, "Brain tumors," *Am. J. Med.*, vol. 131, no. 8, pp. 874–882, 2018.

[2]     E. S. Biratu, F. Schwenker, Y. M. Ayano, and T. G. Debelee, "A survey of brain tumor segmentation and classification algorithms," *J. Imaging*, vol. 7, no. 9, p. 179, 2021.

[3]     M. A. L. Khaniki, N. Mahjourian, and M. Manthouri, "Hierarchical SegNet with Channel and Context Attention for Accurate Lung Segmentation in Chest X-ray Images," *arXiv Prepr. arXiv2405.12318*, 2024.

[4]     P. Sharma and A. P. Shukla, *Brain Tumor Classification Using Convolution Neural Network*, vol. 341. Springer Singapore, 2022. doi: 10.1007/978-981-16-7118-0_50.

[5]     F. Ghazouani, P. Vera, and S. Ruan, "Efficient brain tumor segmentation using swin transformer and enhanced local self-attention," *Int. J. Comput. Assist. Radiol. Surg.*, vol. 19, no. 2, pp. 273–281, 2024.

[6]     H. Ajami, M. K. Nigjeh, and S. E. Umbaugh, "Unsupervised white matter lesion identification in multiple sclerosis ( MS ) using MRI segmentation and pattern classification : a novel approach with CVIPtools," vol. 12674, pp. 1–6, 2023, doi: 10.1117/12.2688268.

[7]     A. Vaswani *et al.*, "Attention is all you need," *Adv. Neural Inf. Process. Syst.*, vol. 30, 2017.

[8]     M. A. L. Khaniki, M. Mirzaeibonehkhater, and M. Manthouri, "Enhancing Pneumonia Detection using Vision Transformer with Dynamic Mapping Re-Attention Mechanism," in *2023 13th International Conference on Computer and Knowledge Engineering*

*(ICCKE)*, IEEE, 2023, pp. 144–149. doi: 10.1109/ICCKE60553.2023.10326313.

[9]  O. Ovadia, A. Kahana, P. Stinis, E. Turkel, D. Givoli, and G. E. Karniadakis, "Vito: Vision transformer-operator," *Comput. Methods Appl. Mech. Eng.*, vol. 428, p. 117109, 2024.

[10] C. Xia, X. Wang, F. Lv, X. Hao, and Y. Shi, "Vit-comer: Vision transformer with convolutional multi-scale feature interaction for dense predictions," in *Proceedings of the IEEE/CVF Conference on Computer Vision and Pattern Recognition*, 2024, pp. 5493–5502.

[11] M.-T. Luong, H. Pham, and C. D. Manning, "Effective approaches to attention-based neural machine translation," *arXiv Prepr. arXiv1508.04025*, 2015.

[12] K. Xu *et al.*, "Show, attend and tell: Neural image caption generation with visual attention," in *International conference on machine learning*, PMLR, 2015, pp. 2048–2057.

[13] M. A. Labbaf Khaniki, M. Mirzaeibonehkhater, A. Samii, and M. Manthouri, "Adaptive Control of Spur Gear Systems via Proximal Policy Optimization and Attention-Based Learning," in *2023 9th International Conference on Control, Instrumentation and Automation (ICCIA)*, 2023, pp. 1–5. doi: 10.1109/ICCIA61416.2023.10506397.

[14] Z. Huang, X. Wang, L. Huang, C. Huang, Y. Wei, and W. Liu, "Ccnet: Criss-cross attention for semantic segmentation," in *Proceedings of the IEEE/CVF international conference on computer vision*, 2019, pp. 603–612.

[15] C. F. Chen, Q. Fan, and R. Panda, "CrossViT: Cross-Attention Multi-Scale Vision Transformer for Image Classification," *Proc. IEEE Int. Conf. Comput. Vis.*, pp. 347–356, 2021, doi: 10.1109/ICCV48922.2021.00041.

[16] Z. Lu and E. Elhamifar, "Fact: Frame-action cross-attention temporal modeling for efficient action segmentation," in *Proceedings of the IEEE/CVF Conference on Computer Vision and Pattern Recognition*, 2024, pp. 18175–18185.

[17] J. Cheng *et al.*, "Enhanced performance of brain tumor classification via tumor region augmentation and partition," *PLoS One*, vol. 10, no. 10, p. e0140381, 2015.

[18]  J. Wang, S.-Y. Lu, S.-H. Wang, and Y.-D. Zhang, "RanMerFormer: Randomized vision transformer with token merging for brain tumor classification," *Neurocomputing*, vol. 573, p. 127216, 2024, doi: https://doi.org/10.1016/j.neucom.2023.127216.

[19]  T. Usuzaki *et al.*, "Grading diffuse glioma based on 2021 WHO grade using self-attention-base deep learning architecture: variable Vision Transformer (vViT)," *Biomed. Signal Process. Control*, vol. 91, p. 106001, 2024, doi: https://doi.org/10.1016/j.bspc.2024.106001.

[20]  M. S. Vahdatpour, "Addressing the Knapsack Challenge Through Cultural Algorithm Optimization," *arXiv Prepr. arXiv2401.03324*, 2023, doi: 10.48550/arXiv.2401.03324.

[21]  A. K. Newendorp, A. J. Perron, M. J. Sells, K. T. Nelson, M. C. Dorneich, and S. B. Gilb, "Apple ' s Knowledge Navigator : Why Doesn ' t that Conversational Agent Exist Yet ?", doi: 10.1145/3613904.3642739.

[22]  M. Kiaghadi, M. Sheikholeslami, A. M. Alinia, and F. M. Boora, "Predicting the performance of a photovoltaic unit via machine learning methods in the existence of finned thermal storage unit," *J. Energy Storage*, vol. 90, p. 111766, 2024, doi: 10.1016/j.est.2024.111766.

[23]  M. Salehi, N. Javadpour, B. Beisner, M. Sanaei, and S. B. Gilbert, "Innovative Cybersickness Detection: Exploring Head Movement Patterns in Virtual Reality," *arXiv Prepr. arXiv2402.02725*, 2024, doi: 10.48550/arXiv.2402.02725.

[24]  M. Sanaei, S. B. Gilbert, N. Javadpour, H. Sabouni, M. C. Dorneich, and J. W. Kelly, "The Correlations of Scene Complexity, Workload, Presence, and Cybersickness in a Task-Based VR Game," *arXiv Prepr. arXiv2403.19019*, 2024, doi: 10.48550/arXiv.2403.19019.

[25]  M. Farhang and F. Safi-Esfahani, "Recognizing mapreduce straggler tasks in big data infrastructures using artificial neural networks," *J. Grid Comput.*, vol. 18, no. 4, pp. 879–901, 2020, doi: 10.1007/s10723-020-09514-2.

[26]  M. S. Vahdatpour and Y. Zhang, "Latency-Based Motion Detection in Spiking Neural Networks," *Int. J. Cogn. Lang. Sci.*, vol. 18, no. 3, pp. 150–155, 2024.

[27] M. Younesi Heravi, I. S. Dola, Y. Jang, and I. Jeong, "Edge AI-Enabled Road Fixture Monitoring System," *Buildings*, vol. 14, no. 5, p. 1220, 2024, doi: 10.3390/buildings14051220.

[28] S. Salahshour, M. Salimi, K. Tehranian, N. Erfanibehrouz, M. Ferrara, and A. Ahmadian, "Deep prediction on financial market sequence for enhancing economic policies," *Decis. Econ. Financ.*, pp. 1–20, 2024.

[29] M. Bigonah, F. Jamshidi, and D. Marghitu, "Immersive Agricultural Education: Gamifying Learning With Augmented Reality and Virtual Reality," in *Cases on Collaborative Experiential Ecological Literacy for Education*, IGI Global, 2024, pp. 26–76.

[30] M. Bigonah, F. Jamshidi, A. Pant, and D. Marghitu, "Work in Progress: Grace Platform: Enhancing Pedagogy with Gamified AR and VR in Agriculture Education," in *2024 ASEE Annual Conference & Exposition*, 2024.

[31] S. Tayebati, T. Tulabandhula, and A. R. Trivedi, "Sense Less, Generate More: Pre-training LiDAR Perception with Masked Autoencoders for Ultra-Efficient 3D Sensing," *arXiv Prepr. arXiv2406.07833*, 2024.

[32] S. Nouraniboosjin, M. Yousefi, S. Meisami, M. Yousefi, and S. Meisami, "Empowering Healthcare: A Blockchain-Based Secure and Decentralized Data Sharing Scheme with Searchable Encryption," *Int. J. Cybern. Informatics*, vol. 13, no. 13, p. 47, 2024.

[33] S. Deepak and P. M. Ameer, "Brain tumor classification using deep CNN features via transfer learning," *Comput. Biol. Med.*, vol. 111, p. 103345, 2019.

[34] M. Mehrnia, M. Elbayumi, and M. S. M. Elbaz, "Assessing Foundational Medical'Segment Anything'(Med-SAM1, Med-SAM2) Deep Learning Models for Left Atrial Segmentation in 3D LGE MRI," *arXiv Prepr. arXiv2411.05963*, 2024.

[35] M. K. Nigjeh, H. Ajami, and S. E. Umbaugh, "Automated classification of white matter lesions in multiple sclerosis patients ' MRI images using gray level enhancement and deep learning," vol. 12674, pp. 1–6, 2023, doi: 10.1117/12.2688269.

[36] M. Y. Heravi, Y. Jang, I. Jeong, and S. Sarkar, "Deep learning-based activity-aware 3D

human motion trajectory prediction in construction," *Expert Syst. Appl.*, vol. 239, p. 122423, 2024, doi: 10.1016/j.eswa.2023.122423.

[37] K. Khanjani, S. R. Hosseini, S. Shashaani, and M. Teshnehlab, "COVID-19 Detection Based on Blood Test Parameters using Various Artificial Intelligence Methods," *arXiv Prepr. arXiv2404.02348*, 2024, doi: 10.48550/arXiv.2404.02348.

[38] V. K. Bairagi, P. P. Gumaste, S. H. Rajput, and K. S. Chethan, "Automatic brain tumor detection using CNN transfer learning approach," *Med. Biol. Eng. Comput.*, vol. 61, no. 7, pp. 1821–1836, 2023.

[39] A. Ashary, M. M. Rayguru, P. SharafianArdakani, I. Kondaurova, and D. O. Popa, "Multi-Joint Adaptive Motion Imitation in Robot-Assisted Physiotherapy with Dynamic Time Warping and Recurrent Neural Networks," in *SoutheastCon 2024*, IEEE, 2024, pp. 1388–1394. doi: 10.1109/SoutheastCon52093.2024.10500261.

[40] S. Kumar and D. Kumar, "Human brain tumor classification and segmentation using CNN," *Multimed. Tools Appl.*, vol. 82, no. 5, pp. 7599–7620, 2023.

[41] M. Aloraini, A. Khan, S. Aladhadh, S. Habib, M. F. Alsharekh, and M. Islam, "Combining the transformer and convolution for effective brain tumor classification using MRI images," *Appl. Sci.*, vol. 13, no. 6, p. 3680, 2023.

[42] G. J. Ferdous, K. A. Sathi, M. A. Hossain, M. M. Hoque, and M. A. A. Dewan, "LCDEiT: A linear complexity data-efficient image transformer for MRI brain tumor classification," *IEEE Access*, vol. 11, pp. 20337–20350, 2023.

[43] M. Mowlai, M. Mahdavimanshadi, and I. Sayyadzadeh, "Adapting Transformer-Based Multi-Style Networks for Human Pose Prediction with a Custom Data Pipeline in Industrial Human-Robot Collaboration," in *2024 Systems and Information Engineering Design Symposium (SIEDS)*, IEEE, 2024, pp. 274–279. doi: 10.1109/SIEDS61124.2024.10534732.

[44] J. Liang, C. Yang, J. Zhong, and X. Ye, "BTSwin-Unet: 3D U-shaped symmetrical Swin transformer-based network for brain tumor segmentation with self-supervised pre-training," *Neural Process. Lett.*, vol. 55, no. 4, pp. 3695–3713, 2023.

[45] J. Lin *et al.*, "CKD-TransBTS: clinical knowledge-driven hybrid transformer with modality-correlated cross-attention for brain tumor segmentation," *IEEE Trans. Med. Imaging*, 2023.

[46] L. Zongren, W. Silamu, F. Shurui, and Y. Guanghui, "Focal cross transformer: Multi-view brain tumor segmentation model based on cross window and focal self-attention," *Front. Neurosci.*, vol. 17, p. 1192867, 2023.

[47] M. Doyle, T. F. Fuller, and J. Newman, "Modeling of galvanostatic charge and discharge of the lithium/polymer/insertion cell," *J. Electrochem. Soc.*, vol. 140, no. 6, p. 1526, 1993.